\documentclass[12pt]{iopart}

\usepackage[pdftex]{graphicx}
\usepackage{iopams}
\usepackage{epstopdf}
\usepackage{graphicx}
\usepackage{cite}
\usepackage{gensymb}

\begin{document}
\def\be{\begin{equation}}
\def\ee{\end{equation}}
\def\bea{\begin{eqnarray}}
\def\eea{\end{eqnarray}}
\def\rp{r_{+}}
\def\rmm{r_{-}}

\title{Geometrothermodynamics of the  Kehagias-Sfetsos  Black Hole }
\date{April 2010}

\author{W.\  Janke}
\address{Institut f\"ur Theoretische Physik and Centre for Theoretical Sciences (NTZ),
 Universit\"at Leipzig,
 Postfach 100\,920,
04009 Leipzig, Germany}

\author{D.\ A.\ Johnston}
\address{Department of Mathematics and the Maxwell Institute for Mathematical
Sciences,
Heriot-Watt University, Riccarton, Edinburgh EH14 4AS, Scotland}

\author{R.\ Kenna}
\address{Applied Mathematics Research Centre, Coventry University, Coventry, CV1 5FB, England}


\begin{abstract}
The application of information geometric ideas  to statistical mechanics using a metric on the space of states, pioneered by Ruppeiner
and Weinhold,
has proved to be a useful alternative approach to characterizing phase transitions. Some puzzling anomalies become apparent, however,
when these methods are applied to the study of black hole thermodynamics. A possible resolution was suggested by Quevedo
{\it et al.} who emphasized the importance of Legendre invariance in thermodynamic metrics. They found physically consistent results for various black holes
when using a Legendre invariant metric, which  agreed with a direct determination of the properties of phase transitions from the specific heat.

Recently, information geometric methods have been employed by Wei  {\it et al.} to study the Kehagias-Sfetsos (KS) black hole in Ho\v{r}ava-Lifshitz  gravity. The formalism 
suggests that a coupling parameter in this theory plays a role analogous to the charge in Reissner-Nordstr\"om  (RN) black holes or angular momentum 
in the Kerr  black hole and calculation of the specific heat 
shows a singularity which may be interpreted as a phase transition.  When the curvature of the Ruppeiner metric is calculated for such a theory it does not, however, show a singularity at the phase transition point. 

We show that the curvature of a particular Legendre invariant (``Quevedo'') metric for the KS black hole {\it is}  singular at the phase transition point. We contrast the results for the Ruppeiner, Weinhold and Quevedo metrics and in the latter case investigate the consistency of taking either the entropy or mass as the thermodynamic potential.

\end{abstract}

\maketitle


\section{Introduction}

The thermodynamics of black holes has been studied extensively
since the work of Hawking \cite{Hawk}. The notion of critical  behaviour has arisen in several
contexts for black holes, ranging from the Hawking-Page \cite{HP}
phase transition in hot anti-de-Sitter space and the pioneering work by
Davies  \cite{Davies} on the thermodynamics of Kerr-Newman black holes, to the
idea that the extremal limit of various black hole families might themselves be regarded
as genuine critical points \cite{Louko, Chamblin, Cai}. As for standard statistical mechanical systems,
critical points are  signalled by singularities in the specific heat. 

More recently, various groups have investigated the application of ideas from information geometry
to the study of black hole thermodynamics. The use of information geometry \cite{Fish} in statistical mechanics 
in general was largely pioneered  by Ruppeiner 
\cite{Ruppeiner1} and Weinhold \cite{weinhold}, who suggested that the curvature of a metric defined on the 
space of parameters of a statistical mechanical theory could provide information about the phase structure.
Specifically, from consideration of fluctuations, Ruppeiner suggested a metric based on the entropy
\begin{eqnarray}
 g^R_{ij}
           =- \partial_{i} \partial_{j} S (M, E^a )
\end{eqnarray}
where $S$ is the entropy, $M$ is the mass and $E^a$ are the other extensive thermodynamic variables of the theory under consideration. It was found that the
curvature of this metric was zero for a non-interacting theory such as an ideal gas, but non-zero
for an interacting theory such as a van der Waals gas, and divergent at the phase transition points
\cite{Ruppeiner2}.

The Ruppeiner metric is conformally related to the Weinhold metric \cite{weinhold} by
\be{
\label{eq:conformal}
g^W_{ij} = T g^R_{ij},
}
\ee
where $T$ is the temperature of the system under consideration. This Weinhold metric is defined as
the Hessian of the energy (mass) with respect to entropy and other extensive parameters, namely
\be{
g^W_{ij} = \partial_i \partial_j M (S, E^a) \; .
}
\ee
For non-black-hole systems, the results from using either metric have proved to be consistent \cite{Jany, Brody, Brian, no1, no2, no3},
but consideration of different black hole families under various assumptions has led to numerous puzzling 
results for both metrics and inconsistencies between them \cite{Aman1,thesis,Aman2,Aman3,Aman4,Shen,Medved,Chakraborty,Sarkar,Mirza}. 

Wei {\it et al.} \cite{Wei} have recently added to this catalogue of inconsistencies by examining the Ruppeiner metric and curvature for the  Kehagias-Sfetsos (KS) black hole in Ho\v{r}ava-Lifschitz gravity (HL) and finding no singularity in the curvature 
at a point where a direct calculation of the specific heat does indicate a singularity. The Legendre invariant metric suggested by Quevedo {\it et.al.}
\cite{quev07, quev080, quev08, quev08a, quev09, quev09a} has proved more successful in capturing the phase structure of other black hole families than the Ruppeiner metric,
and in this note we calculate the Quevedo metric and curvature for the KS black hole using both the entropy and mass
as the thermodynamic potential.

In the sequel we first briefly define the action for Ho\v{r}ava-Lifschitz gravity and sketch the KS black hole solution. We then describe the Ruppeiner metric
for the  KS black holes, before moving on to the Quevedo metric  in both the entropy and mass
representation. An explicit expression $M(S,P)$ for the KS black hole is presented, which is useful in calculating the metric
and basic thermodynamic quantities. In conclusion,
the general features of  the scaling of the Quevedo curvature and their origin
are highlighted.


\section{Ho\v{r}ava-Lifschitz gravity}

The suggestion by Ho\v{r}ava \cite{H1,H2,H3} that an anisotropic theory of gravity at a Lifshitz point \cite{L1}  might offer a viable quantum field theory
of gravity while still retaining the properties of Einstein gravity in the IR has led to an explosion of recent work. Since the theory breaks general covariance
to 3D spatial covariance plus time re-parametrization invariance it is most naturally couched in the (3+1) language of the ADM  \cite{ADM} formalism, where a general metric is written as
\begin{equation}
ds^{2}=-N^{2}dt^{2}+g_{ij}\left(dx^{i}+N^{i}dt\right)\left(dx^{j}+N^{j}dt\right).
\end{equation}
The lapse and shift can then be used to construct the extrinsic curvature of the 3-space
\begin{equation}
K_{ij}=\frac{1}{2N}\left(\dot{g}_{ij}-\nabla_{i}N_{j}-\nabla_{j}N_{i}\right),
\end{equation}
where the $\dot{g}_{ij}$ is the time derivative of the metric on the spatial slice.
The Ho\v{r}ava-Lifshitz  action may then be written as
\begin{eqnarray}
S_{HL} &=& \int dtdx^{i}\,
\sqrt{g}N\left(\mathcal{L}_{0}+\tilde{\mathcal{L}}_{1}\right), 
\end{eqnarray}
with
\begin{eqnarray}
\mathcal{L}_{0} &=& \frac{2}{\kappa^{2}}\left(K_{ij}K^{ij}-\lambda K^{2}\right)
                   +\frac{\kappa^{2}\mu^{2}\left(\Lambda_{W}R^{(3)}-3\Lambda_{W}^{2}\right)}
                          {8(1-3\lambda)}\,, \label{L0} \\
\tilde{\mathcal{L}}_{1}&=&\frac{\kappa^{2}\mu^{2}(1-4\lambda)}{32(1-3\lambda)}\left(R^{(3)}\right)^{2}
                        -\frac{\kappa^{2}}{2w^{4}}\left(C^{(3)}_{ij}-\frac{\mu w^{2}}{2}R^{(3)}_{ij}\right)
                   \left(C^{(3)ij}-\frac{\mu
                   w^{2}}{2}R^{(3)ij}\right) \nonumber
\end{eqnarray}
where $\Lambda_{W}$, $\kappa$, $\lambda$, $\mu$ and $\omega$ are various constants and $R^{(3)}_{ij}$ and $R^{(3)}$
are the three-dimensional Ricci tensor and Ricci scalar. The Cotton tensor for the three-geometry, which also appears, is defined as
\begin{eqnarray}
 C^{(3)ij}=\epsilon^{ijk}\nabla_{k}\bigg(R_{\,l}^{(3)j}
         -\frac{1}{4}R^{(3)}\delta_{\,l}^{j}\bigg).
\end{eqnarray}
As it stands the generic IR vacuum of such a theory is anti-de Sitter, but it is possible to deform the theory
with an additional relevant operator $\mu^{4}R^{(3)}$, which allows a Minkowski vacuum \cite{KS}. Using $x^0 = ct$
the IR limit of this augmented action matches the Einstein-Hilbert action 
\begin{equation}
 S_{EH}=\frac{1}{16\pi G}\int d^{4}x\,
        N\sqrt{g}(K_{ij}K^{ij}-K^{2} + R^{(3)})
\end{equation}
in the limit $\Lambda_{W}\rightarrow0$ and $\lambda=1$ if
\begin{equation}
 c^{2}=\frac{\kappa^{2}\mu^{4}}{2},\;\;G=\frac{\kappa^{2}}{32\pi c},
\end{equation}
The augmented action with 
$\mu^{4} R^{(3)}$  considered in the limit
$\Lambda_{W}\rightarrow0$ is usually denoted  ``deformed HL
gravity" \cite{bMyung}. 


\section{The KS black hole solution}

In \cite{KS} Kehagias and Sfetsos showed that
the deformed HL gravity at $\lambda=1$ admits a Schwarzschild-like  black hole solution, where a metric ansatz  
\begin{eqnarray}
 ds_{HL}^2 = - N^2(r)\,dt^2 + \frac{1}{f(r)}dr^2 + r^2 (d\theta^2
             +\sin^2\theta d\phi^2).
\end{eqnarray}
leads to 
\begin{eqnarray}
 N^2=f=1 + \omega r^2-\sqrt{r(\omega^2 r^3+4\omega M)} \; .
 \label{solution}
\end{eqnarray}
$M$ is an integration constant which is related to the mass of the black hole,
as can be seen by noting that 
\begin{equation}
	f \approx 1 - \frac{2 M}{r} + \mathcal{O} (r^{-4} )
\end{equation}
when $r \gg (2 M / \omega)^{1/3}$ which is the standard Schwarzschild behaviour. The KS black hole displays two event horizons
at
\begin{equation}
	r_{\pm} =  M \pm \sqrt{M^2 - \frac{1}{2 \omega}},
\end{equation}
which is strikingly similar to the formula giving the event horizons in the
Reissner-Nordstr\"om (RN) black hole in standard Einstein gravity
\begin{equation}
	r_{\pm} =  M \pm \sqrt{M^2 - Q^2} ,  \label{rRN}
\end{equation}
or that for the Kerr black hole
\begin{equation}
	r_{\pm} =  M \pm \sqrt{M^2 - \frac{J^2}{M^2}}. \label{rK}
\end{equation}
This has led to the suggestion that 
$P=\sqrt{\frac{1}{2\omega}}$ should be treated as a charge-like parameter when considering the thermodynamics of the KS black hole
\cite{Peng,Myung,Wang}. If one does this, the mass $M$, Hawking temperature $T$ and specific heat $C$  for the KS black hole may be written 
in a similar manner to those for the RN and Kerr black holes giving  \cite{Wei}
\begin{eqnarray} \label{CKS}
M &=& \frac{r_{+}  + r_{-}}{2}  \; ,\nonumber \\
T &=&  \frac{r_{+}-r_{-}}{4\pi r_{+}(r_{+}+2r_{-})} \; , \\
C &=& - \frac{2 \pi r_{+}(r_{+}+2r_{-})^{2}(r_{+}-r_{-})}
              {r_{+}^{2} - 5r_{+}r_{-}-2r_{-}^{2}}. \nonumber
\end{eqnarray}

The entropy may also be calculated 
and written in a Beckenstein-Smarr \cite{BS1,BS2} like manner, giving \cite{cai1,cai2,cai3,Castillo}
\begin{eqnarray} \label{SKS}
S  &=&  \pi ( M + \sqrt{M^2 - P^2} )^2 + 4 \pi P^2 \ln ( M + \sqrt{M^2 - P^2} ) + S_0 \nonumber \\
       &=&  \pi r_{+}^2  + 4 \pi r_{+} r_{-} \ln \left( r_{+} \right)  + S_0  \; .
\end{eqnarray}
$S_0$
is a constant of integration which plays no role in the sequel. It is natural to set $S_0=0$ to match 
up with the Schwarzschild limit.

The singularity in the specific heat can be interpreted as signalling a phase transition
at $5r_{+}r_{-}-r_{+}^{2}+2r_{-}^{2}=0$, i.e. 
$r_{+} = [5/2 +\sqrt{33}/2]\,  r_{-}\, \, $,
for the KS black hole.


\section{Ruppeiner information geometry of the KS black hole}

The Ruppeiner metric components for the KS black hole were calculated by Wei {\it et al.} \cite{Wei}
using equ.(\ref{SKS})
\begin{eqnarray}
 g^{R}_{11}&=&-\frac{8\pi r_{+}(r_{+}^2-5r_{-}r_{+}-2r_{-}^2)}{(r_{+}-r_{-})^{3}},\nonumber\\
 g^{R}_{12}&=&g^{R}_{21}=-\frac{16\pi ( r_+ r_- )^{1/2} (r_+^2+r_+r_-+r_-^2)}{(r_+-r_-)^3},\\
 g^{R}_{22}&=&\frac{4\pi(r_+^3 + 10 r_+^2r_ - - 5 r_+ r_-^2 + 6 r_-^3)}{(r_+-r_-)^3}
            -8\pi\ln(r_+).\nonumber
\end{eqnarray}
The Ricci scalar for the Ruppeiner metric is then found to be 
\begin{eqnarray}
 R^{R}=\frac{(r_+ + 2 r_- ) (r_+^2 + 7 r_+ r_- + r_-^2)}{\pi r_+\left[
   r_+^2 + 16 r_+ r_- +  4 r_-^2- 2 (r_+^2 - 5 r_+
   r_- - 2 r_-^2) \ln r_+\right]^2}
\end{eqnarray}
but it fails to show a singularity at the point, $ r_{+}^2- 5 r_{+} r_{-} - 2 r_{-}^2 = 0
$, where the specific heat has a singularity in equ.(\ref{CKS}), although the pre-factor of the log term in the denominator does vanish
at this point. In addition, it is neither zero nor singular in the extremal limit $r_+ \to r_-$.

The KS black hole thus adds a further example to the (long) list of peculiarities which arise when information geometry, in the form of the Ruppeiner or Weinhold metric, is applied to the thermodynamics of black holes.
A possible resolution of such difficulties in general in the context of black hole thermodynamics  was suggested by Quevedo {\it  et al.} \cite{quev07,quev080, quev08, quev08a, quev09a,quev09}. They argued that an important feature for thermodynamic metrics was Legendre invariance, which
was not a property of either the Ruppeiner or Weinhold metrics. They found consistent results for various black holes
when using a Legendre invariant metric definition, which agreed with direct calculations of phase transition points from the specific heat. In the next section we apply the formalism to the KS black hole. 


\section{Geometrothermodynamics of the KS black hole in the entropy representation}

Quevedo  {\it et al.}'s starting point \cite{quev07} was the observation that standard thermodynamics was invariant
with respect to Legendre transformations, since one expects consistent results whatever starting potential one takes,
and they coined the name geometrothermodynamics for a formalism which ensured this. Their work was based on
the use of contact geometry as a framework for thermodynamics, developed by Hermann \cite{Hermann}, Mruga{\l}a \cite{Mrugala} and others.

For the geometrothermodynamics  of black holes they considered a $2 n +1$ dimensional thermodynamic 
phase space ${\cal T}$ with independent coordinates 
$\{\Phi, E^a,I^a\}$, $a=1,...,n$, where $\Phi$ represents the thermodynamic potential, and $E^a$ and $I^a$ are the 
extensive and intensive thermodynamic variables, respectively. This thermodynamic  phase space was
endowed with a Gibbs one-form 
$\Theta = d\Phi - \delta_{ab} I^a d E ^b$, $\delta_{ab} = {\rm diag}(1,...,1)$,
 and the Legendre transform invariant metric
\bea \label{QG}
G &=& (d\Phi - \delta_{ab} I^a d E^b)^2 +  (\delta_{ab}E^a I^b)(\eta_{cd} dE^c dI^d) \, , \nonumber \\
 \eta_{cd} &=& {\rm diag}(-1,1,...,1) \ ,
\eea
which was invariant with respect to  $ \{\Phi, E^a,I^a\}\rightarrow \{\tilde \Phi, \tilde E ^a, \tilde I ^ a\}$,
with 
$ \Phi = \tilde \Phi - \delta_{ab} \tilde E ^a \tilde I ^b \ ,\ 
 E^a = - \tilde I ^ {a}, \  I^{a} = \tilde E ^ a $. The Gibbs one-form satisfies the condition
$\Theta \wedge (d \Theta )^n \ne 0$, making it a contact form and the triplet $\{ {\cal T}, \Theta, G \}$ constitutes a Riemannian
contact manifold.

The equilibrium space ${\cal E} \subset {\cal T}$  is then defined by $\varphi: \{E^a\}  \mapsto \{\Phi, E^a, I^a\}$, satisfying the condition 
$\varphi^* (\Theta) =0 $. This means that  on ${\cal E}$ the first law of thermodynamics holds, $d\Phi = \delta_{ab} I^a d E ^b$, and the 
equilibrium conditions $I^a = \delta^{ab}\partial \Phi/\partial E^b$ give the $I^a$ in terms of  the $E^a$. The induced thermodynamic metric
on  ${\cal E}$, which plays a similar role to the Ruppeiner or Weinhold metric and which we denote here as the Quevedo metric, is given by
\be
g^Q=\left(E^c\frac{\partial \Phi}{\partial E^c}\right)
\left(\eta_{ab}\delta^{bc}\frac{\partial^2\Phi}
{\partial E^c \partial E^d}
dE^a dE^d\right) \ .
\label{gQ}
\ee 
The choice of $\eta_{cd}$ in equ.(\ref{QG}) rather than $\delta_{cd}$, which is also possible, prevents off diagonal terms $g_{1k}, \; k \ne 1$, appearing which in turn plays a vital role in determining the singularities of the curvature.

In the case of the KS 
black hole using the entropy as the thermodynamic potential one considers the 5-dimensional thermodynamic phase space ${\cal T}$  with coordinates $Z^a = \{S, E^a, I^a\}=\{S, M, P, 1/T, -V_P/T\}$.
The fundamental one-form
in  this $S-$representation is given by 
\be
\Theta_S =  dS -\frac{1}{T} dM  +\frac{V_P}{T} d P  \  ,
\ee
so defining the space of equilibrium states ${\cal E}$ by $\varphi^*_S(\Theta_S) =0$,
generates both the first law of thermodynamics of the KS black hole
\be 
dM = T dS +  V_P d P  \ ,
\label{first}
\ee 
and the equilibrium conditions 
\be
\frac{1}{T} = \frac{\partial  S}{\partial M} \ ,\quad
\frac{V_P}{T} = -\frac{\partial  S}{\partial P} \,\quad
  \ . 
\ee
From equ.(\ref{gQ}) the Quevedo metric in this case is 
\be
g^{Q} = \left(M S_M +P S_P \right)\left( - S_{MM} d M^2 + S_{PP} dP^2  \right) \ ,
\label{gRNQ/}
\ee
which may be written in components as  
\begin{eqnarray}
 g^{Q}_{11}&=& - \frac{16 \pi^2r_+^2 [ r_+ + 2 r_- + 4 r_- \ln (r_+ )](r_+^2 - 5 r_+ r_- -2 r_-^2)}{(r_+ - r_-)^3},\nonumber\\
 g^{Q}_{12}&=&g^{Q}_{21}= 0,\nonumber\\
 g^{Q}_{22}&=& - \frac{8 \pi^2 r_+ [ r_+ + 2 r_- + 4 r_- \ln (r_+ ) ] A(r_+, r_-)}{(r_+ - r_-)^3} ,
\end{eqnarray}
from which the Ricci scalar may be calculated without further ado to give
\bea \label{RKSQ}
R^{Q} &=& \frac{(r_+ - r_-)^2 }{2 
\pi^2 r_+^2 [r_+ + 2 r_- + 4 r_- \ln (r_+ ) ]^3 } \times\nonumber \\
&{}& \frac{B(r_+, r_-)}{(r_+^2 - 5 r_+ r_- -2 r_-^2)^2 \, A( r_+, r_-)^2}
\eea
where 
\bea
A ( r_+, r_-) &=& r_+^3  +  10 r_+^2 r_-  - 5 r_+ r_-^2 + 6 r_-^3  - 2 ( r_+ - r_-)^3 \ln (r_+ ) 
\eea
and $B ( r_+, r_-) $ is  a long and not very illuminating expression which is neither zero nor divergent
when $r_+ = r_-$ and $r_+^2 - 5 r_+ r_- -2 r_-^2=0$ (i.e. $r_+ = [ 5/2 +\sqrt{33}/2] \, r_-$).
We give it in the appendix for completeness.
Looking at  the curvature $R^{Q}$
we see that it  
diverges as $\Delta^{-2}$, with $\Delta = (r_+^2 - 5 r_+ r_- -2 r_-^2)$, at the same point as the specific heat. 

The specific heat vanishes in the extremal limit, $r_-  \to  r_+$ since $C = T ( \partial S / \partial T)$ and $T \to 0$ in this limit. 
This is also the case for the curvature in equ.(\ref{RKSQ}) calculated in the entropy representation, but this is 
not a generic feature, as we see in the next section by calculating the Quevedo curvature using the mass
as the thermodynamic potential. 


\section{Geometrothermodynamics  of the KS black hole in the mass representation}

There is a degree of arbitrariness in the definition of the metric $G$ on $\cal{T}$ (and hence $g^Q$), since various choices will 
allow for the desired Legendre invariance. Different choices for the thermodynamic potential $\Phi$ are also possible. In the preceding section we have taken $\Phi = S(M,P)$, but $\Phi = M(S,P)$ would have been {\it a priori} equally valid.  Some of the properties
such as the relation between the specific heat and curvature singularities of the Quevedo metric are particularly apparent in this mass representation
\be
g^{MQ} = \left(S M_S +P M_P \right)\left( -M_{SS} d S^2 + M_{PP} dP^2  \right) \ ,
\label{gMQ}
\ee
where, at the risk of overcomplicating the notation, we write the superscript $MQ$ to denote the
use of the mass $M$ as the thermodynamic potential.  
In these variables the specific heat may be evaluated as $C = M_{S}/M_{SS}$ and the presence of $\eta_{cd}$ in the definition of $G, g^{MQ}$ ensures
the absence of off-diagonal terms, $\partial^2 M / \partial S \partial P$ and 
a $(M_{SS})^{-2}$ factor in the resulting curvature. 
 
For the RN and Kerr black holes it is straightforward to express $M$ as a function of $S,Q$ or $S,J$ using the Beckenstein-Smarr formulae 
\bea \label{MSQ}
M_{RN}(S,Q) =  \frac{ S + \pi Q^2}{ ( 4 \pi S)^{1/2}} \nonumber \\
M_{\rm{Kerr}}(S,J) =  \left( \frac{ S^2 + 4 \pi^2 J^2}{4 \pi S} \right)^{1/2}  \, .
\eea
It is more difficult to write $M(S,P)$ explicitly for the KS black hole using equ.(\ref{SKS}), but this may still be done using the Lambert W function, which is the solution
of 
\be
W(x) \cdot \exp ( W(x)) =x ,
\ee
to give
\be 
M_{KS} (S,P) = \frac{ P \left[ 1 + 2 W \left( \frac{ \exp ( S / 2 \pi P^2  ) }{2 P^2} \right) \right] } { 2^{3/2} \left[ W \left(\frac{ \exp ( S / 2 \pi P^2 )}{2 P^2} \right) \right]^{1/2}} \, , \label{MW}
\ee
where we have assumed that $S_0 = 0$. Choosing a non-zero constant simply shifts $S \to S - S_0$.

The specific heat calculated from $C = M_S / M_{SS}$ in these variables 
for the KS black hole is
\be 
C = \frac{- 4  P^2 \pi  \left[ 2 W \left( \frac{ \exp ( S / 2 \pi P^2  ) }{2 P^2} \right) -1 \right] \left[ W \left( \frac{ \exp ( S / 2 \pi P^2  ) }{2 P^2} \right) + 1 \right]^2 } { \left[2  W^2 \left(\frac{ \exp ( S / 2 \pi P^2 )}{2 P^2} \right) 
- 1 - 5 W \left(\frac{ \exp ( S / 2 \pi P^2 )}{2 P^2} \right) 
\right]} \, . \label{CKSW}
\ee
Rewriting this in terms of  $r_{\pm}$'s reproduces (as it should) the expression for $C$ given in equ.(\ref{CKS}), which had been calculated previously \cite{Wei} using
\be
C = \frac{\frac{\partial M}{\partial r_+}\bigg|_{P} }{\frac{\partial T}{\partial r_+}\bigg|_{P}} \, .
\ee
The additional observation from solving equ.(\ref{MW}) for $W$ in terms of $r_{\pm}$ that
$W$ is equal to $r_{+} / 2 r_{-}$ is useful for rewriting equ.(\ref{CKSW}).

The curvature of the Quevedo metric in the mass representation for the KS black hole
is a rather cumbersome expression , but it takes the form
\be
R^{MQ} = \frac{C(S,P)}{D(S,P) \left[2  W^2 \left(\frac{ \exp ( S / 2 \pi P^2 )}{2 P^2} \right) 
- 1 - 5 W \left(\frac{ \exp ( S / 2 \pi P^2 )}{2 P^2} \right) 
\right]^2}
\ee
where $C(S,P), D(S,P)$ are neither singular nor zero at the zeros of the other factor in the denominator.
This  still shows clearly the correspondence between the specific heat and curvature singularities that is expected on general grounds, since substituting $W = r_{+} / 2 r_{-}$ in the denominator 
recovers the same singular factor $[ r_+^2 - 5 r_+ r_- -2 r_-^2]^2$
seen when the entropy is used as the thermodynamic potential.

The behaviour of the Quevedo metric for the KS black hole is thus identical to that of the RN and Kerr
black holes: the location of the singularities of the curvature match those of the specific heat
in both the entropy and mass representations.

\section{The Weinhold geometry of the KS black hole}

With an explicit mass formula, equ.(\ref{MW}), in hand it is also a straightforward matter to calculate the components
of the Weinhold metric for the KS black hole for comparison purposes with both the Ruppeiner and Quevedo metrics. Other black hole families display inconsistencies between the specific heat singularities
and those of the Ruppeiner and Weinhold curvatures. For example, with the RN black hole the Ruppeiner geometry is flat, whereas the Weinhold geometry is curved. 

The KS black hole is no exception to this behaviour.
The metric components from $g^W_{ij} = \partial_i \partial_j M (S, P)$ are found to be
\begin{eqnarray}
 g^{W}_{11}&=& - \frac{( r_+^2 - 5 r_+ r_-  - 2 r_-^2 )}{8 \pi^2 r_+^2 ( r_+ + 2 r_-)^3},\nonumber\\
 g^{W}_{12}&=&g_{21}^{W} = \, \frac{ r_-^{1/2} [\,2  (r_+^2 - 5 r_+ r_-  - 2 r_-^2) \ln ( r_+) - 3 r_+ (r_+ + 2 r_-) \, ]}{2 \pi r_+^{3/2} ( r_+ + 2 r_-)^3},\nonumber\\
 g^{W}_{22}&=& -\frac{1}{r_+ } + \frac{2 (r_+^2 - 11 r_+ r_-  - 2 r_-^2) \ln (r_+ )}{r_+ ( r_+ + 2 r_-)^2 } \nonumber \\
&+& \frac{8 r_- ( r_+^2 - 5 r_+ r_-  - 2 r_-^2) \ln^2 ( r_+ )}{r_+ ( r_+ + 2 r_-)^3 } ,
\end{eqnarray}
which gives a curvature
\bea
R^{W} &=& \frac{ E(r_+ , r_- )}{(r_+ - r_-)^2 [ r_+^2  + 16 r_+ r_- + 4 r_-^2 - 2 ( r_+^2 - 5 r_+ r_-  - 2 r_-^2) \ln (r_+ ) \, ]^2} \nonumber \\
&{}&
\eea
where $E(r_+ , r_- )$ is another complicated expression with no interesting behaviour 
at $r_+^2 - 5 r_+ r_- -2 r_-^2=0$ or $r_+ = r_-$.

$R^{W}$ does {\it not} diverge at the same point as the specific heat, $r_+^2 - 5 r_+ r_- -2 r_-^2=0$, but rather at the extremal limit. Interestingly,
the factor in the denominator is identical to that appearing in the curvature of the Ruppeiner metric, so the pre-factor of the $\ln ( r_+)$ disappears at that point. As for the Ruppeiner metric, the Weinhold metric
does not reproduce the singular behaviour of the KS black hole specific heat.  

\section{Conclusions}

The KS black hole provides a further example of a system in which a particular choice of Legendre invariant Quevedo metric  
captures the phase structure in a manner which eludes the apparently physically well-motivated Ruppeiner 
and Weinhold metrics. It remains a puzzle 
as to why this behaviour manifests itself in  various black hole models  and is not apparent when the Ruppeiner and Weinhold
metrics are used in the description of other, less esoteric, statistical mechanical systems. In the latter case, the use of a Legendre invariant metric does 
not appear to be obligatory to find physically sensible results. 
 
The relation between the singularities of the specific heat and the thermodynamic curvature calculated with this Quevedo metric is consistent
for the black holes and choices of thermodynamic potential discussed here.  
If one accepts the premise that
phase transitions appear as curvature singularities in the thermodynamic metric,
none of the examples examined here and elsewhere using this particular definition have given rise to the sorts of inconsistencies  which have dogged the application of Ruppeiner
and Weinhold metrics in this field. 

The relation between the specific heat and thermodynamic curvature scaling is different to that seen for 
a continuous transition and the Ruppeiner or Weinhold metrics in standard statistical mechanical systems.
In such transitions the Ruppeiner curvature would be  expected to diverge
as the correlation volume, $R \sim \xi^d$, where $\xi$ is some appropriate correlation length.
If the standard scaling assumption $\xi \sim t^{-\nu}$ holds, where $ t = | t - t_c| \to 0$ at the critical point, $R
\sim t^{- \nu d}$. If, in addition, hyperscaling $\nu d = 2 - \alpha$
is also valid, we find $R \sim t^{\alpha -2}$, which relates the singularity of the specific heat $C \sim t^{-\alpha}$
to the singularity of the curvature. As we have seen, the specific heat for the black holes behaves as $C \sim \Delta^{-1}$, but 
the thermodynamic curvature behaves as $\Delta^{-2}$ rather than $\Delta^{-1}$. 

The picture which emerges from the use of a Quevedo metric to investigate the thermodynamics of the KS black hole is thus that, provided $P = (1 / 2 \omega)^{1/2}$ is treated as a charge, both the specific heat and the curvature from the Quevedo metric possess singularities
when $r_+^2 - 5 r_+ r_- -2 r_-^2=0$. The general behaviour is similar to other two parameter families such as the RN and Kerr black holes, which is perhaps not so surprising 
given the similarities in the various thermodynamic formulae for the mass, charges and entropy. This consistency might also be taken as further evidence that the choice of $P$ as a charge-like variable and the resulting expression for $S$ containing a logarithmic term in equ.(\ref{SKS})
is the correct way to reconcile the integral and differential forms 
\be
dM = T dS + V_P dP
\ee
of the first law of thermodynamics in the KS black hole \cite{Wang}.

As a final comment, we emphasize that that the Quevedo metric in the form used here recovers the singularities seen in the specific heat because certain choices (in particular, no off-diagonal elements $g_{1k}$) have been made. It would be interesting to explore whether other choices might also be consistent, and what their physical motivation might be. 

\section{Acknowledgements}

DAJ would like to thank H. Quevedo for some useful clarifications.


\section{Appendix}
The unilluminating factor $B(\rp, \rmm)$ in the numerator of $R^{KS,Q}$ in equ.(\ref{RKSQ}):
\bea
B(\rp, \rmm) = 3 \rp^9 + 66 \rp^8 \rmm +317 \rp^7 \rmm^2 - 506 \rp^6 \rmm^3 -3069 \rp^5 \rmm^4 \nonumber \\ 
- 5110 \rp^4 \rmm^5 + 5221 \rp^3 \rmm^6 +114 \rp^2 \rmm^7 +284 \rp \rmm^8 + 88 \rmm^9 \nonumber \\
{}
\nonumber \\
- 4 \ln ( \rp) \, [2 \rp^9 + 15 \rp^8 \rmm - 78 \rp^7 \rmm^2 -141 \rp^6 \rmm^3 -558 \rp^5 \rmm^4  \nonumber \\
+ 3507 \rp^4 \rmm^5  -3402 \rp^3 \rmm^6 - 1293 \rp^2 \rmm^7 -648 \rp \rmm^8 + 4 \rmm^9] \nonumber \\
{}
\nonumber \\
+ 8 \ln ( \rp)^2 \, ( \rp - \rmm) \,  [\rp^8 - 3 \rp^7 \rmm + 4 \rp^6 \rmm^2 -306 \rp^5 \rmm^3  
+773  \rp^4 \rmm^4 
\nonumber \\
-1227  \rp^3 \rmm^5 -694  \rp^2 \rmm^6 -284 \rp \rmm^7 + 8 \rmm^8         ]
\nonumber \\
{}
\nonumber \\
- 32 \ln ( \rp)^3 \, ( \rp - \rmm)^2 \rp \rmm \, [\rp^5 -12 \rp^4 \rmm +38 \rp^3 \rmm^2  -64 \rp^2 \rmm^3
\nonumber \\
- 51 \rp \rmm^4  -20 \rmm^5]
\eea


\end{document}